\documentclass{optica-article}

\journal{opticajournal} 

\articletype{Research Article}

\usepackage[american]{babel} 
\usepackage[utf8]{inputenc}
\usepackage{indentfirst}
\usepackage{graphicx}
\usepackage{graphics}
\usepackage{float}
\usepackage{fancyhdr}
\usepackage{pgf}
\usepackage{tikz}
\usepackage{blindtext}
\usepackage{amsmath}
\usepackage{physics}
\usepackage[dvipsnames]{xcolor}
\usepackage{cite}
\usepackage{tgbonum}
\usepackage{cancel}
\usepackage{overpic}
\usepackage{pgf}
\usepackage{amsfonts}
\usepackage{pgfplots}
\pgfplotsset{compat=1.18}
\usepackage{amsfonts}

\usepackage{orcidlink}

\makeatletter
\@ifundefined{mathdefault}{}{}
\makeatother

\usetikzlibrary{arrows.meta}
\usepackage{pdfpages}
\usepackage{cleveref}

\usepackage{lineno}

\begin{document}


\title{Deleterious effect of photon-phonon coupling on microcavities in their application as quantum sources.}

\author{Y. Sacha C. L. S.,\authormark{1,*} G. C. Rickli,\authormark{1} J. Dipold,\authormark{1} R. A. Kögler,\authormark{2}  Paulo Nussenzveig,\authormark{1} Nathália B. Tomazio\authormark{1} and M. Martinelli\authormark{1}}


\address{\authormark{1}Instituto de Física, Universidade de São Paulo, Caixa Postal 66318, São Paulo, São Paulo 05315-970, Brazil\\
\authormark{2} Institut für Physik, Humboldt-Universität zu Berlin, Newtonstraße 15, 12489 Berlin, Germany
}

\email{\authormark{*}yurisacha.silva@usp.br} 


\begin{abstract*} 
In quantum systems, the contact with the environment is detrimental to the purity of the state, thus limiting the practical use of entangled sources in quantum information applications. This loss of purity is observed in the form of additional noise in the tomography of the state. 
We investigate this noise dependence in $Si_3N_4$ micro-cavities, previously used for quantum state generation, and demonstrate that the dependence of this noise on the temperature is compatible with a coupling of the photonic chips to a thermal reservoir. 
The control of this noise source is a necessary condition for the efficient implementation of these devices as sources of entangled states in quantum networks.

\end{abstract*}

\section{Introduction}



Entanglement is the fundamental resource for quantum information processing\cite{rspa.2002.1097}. Control over quantum correlations beyond the classical limit and its use in quantum algorithms \cite{10.1145/237814.237866} and quantum communications, either in teleportation \cite{bennett1993teleporting} or quantum key distribution protocols \cite{bennett2014quantum}, ask for robust integrated sources of entangled states for the future of quantum technologies.

Quantum information can be stored and processed in various physical media, including electrons, ions, and photons.
Unfortunately, all of these media face decoherence of the quantum decoherence due to environmental coupling \cite{PhysRevA.44.5401} as an intrinsic bottleneck for practical implementation, leading to the introduction of noise in the system that may eventually degrade the superposition and entanglement to the level of their classical limit. 
We may consider that light offers distinct advantages with respect to losses and decoherence. It couples weakly with its environment and enables rapid propagation, allowing current fiber-optic tools to transport it while preserving the encoded information.

Finally, just as classical computing reduced the physical size of its circuits while significantly increasing their capabilities, quantum computing aims to achieve a substantial increase in computational complexity and processing capacity, alongside a reduction in the physical scale of its systems. An important step towards this goal is to build and characterize a robust and scalable source of entangled states \cite{rspa.2002.1097, PhysRevResearch.2.033316}.

Silicon nitride (Si$_3$N$_4$) photonic chips represent a promising platform for an entangled-state source.
It offers strategic advantages because it can be lithographically patterned using existing technologies for Complementary Metal-Oxide-Semiconductor (CMOS) chip fabrication. It also exhibits a broad transparency range, with very low loss from the visible to the near-infrared, and high optical power tolerance, ensuring a long-lasting, stable platform for a wide range of applications. It was already demonstrated that Si$_3$N$_4$ micro-Kerr resonators above the oscillation threshold can create quantum correlated twin beams
\cite{PhysRevApplied.3.044005,Shen:25} and multipartite quantum correlations in the amplitudes of the fields in frequency combs \cite{cc69-5gq2}. However, previous research by our group found that the beam pairs generated by this source at room temperature do not exhibit entanglement between them due to excess noise in the phase quadratures of these beams
\cite{Alfredo.24}.


There are works in the field that predict increased noise in Si$_3$N$_4$ photonic chips, attributing it to the thermo-refractive properties of the material \cite{PhysRevA.99.061801}. On the other hand, previous research on optical parametric oscillators (OPOs) based on KTP (potassium titanyl phosphate) crystals has already shown that excess noise on the phase of the intense fields can be explained by phonon-photon coupling in the crystal lattice \cite{PhysRevA.79.063816}.
The central objective of this work is to quantitatively characterize the origin and magnitude of this noise.
We demonstrate that the noise is compatible with this last model, thereby coupling the optical system to a thermal bath. This is consistent with the separability of the correlated quantum states generated in \cite{Alfredo.24} and with the observed degradation in squeezing.
By injecting a coherent state into a microring cavity, we could observe an increase in the noise, depending on the temperature $T$ of the chip and on the driving field power $P$, where the coupling rate is estimated in the range of $5.97(12)\times10^{-3}$ $TP$/(WK), in units of standard quantum level.

This paper is organized into four main parts: a mathematical framework to derive the coupling term under consideration, a methodological part where we present the experimental setup alongside the measurement techniques, a section where we focus on the interferometric method in which we determine the intrinsic losses of the micro-resonators, and the results and analysis of the obtained data.

\section{Mathematical Framework}\label{Mat_Frame}

In order to describe the observed excess noise, we review the model proposed in \cite{PhysRevA.79.063816}. Starting from the fluctuations of the electromagnetic field due to permittivity fluctuations, we use the input-output formalism \cite{walls2008quantum} to get a Langevin equation for the time evolution inside the resonator, into which a stochastic term representing the fluctuations in the index of refraction is added, assuming that the phonon reservoir state is unaffected by the electromagnetic field itself. The Brillouin scattering model describes the coupling between the photon and phonon fields \cite{yariv1989quantum}, which is mediated by fluctuations in the refractive index under the reservoir assumption, leading to additional fluctuations in the electromagnetic field.

Consider that the permittivity of the material is described as an average value $\bar\epsilon$ plus a local fluctuation $\delta\epsilon (\vb r,t)$. 
The field is described as a slow varying amplitude $A$ multiplied by a fast oscillation (at optical frequencies)
\begin{align}
    E(\vb r, t) = \operatorname{Re}\left[A(\vb r, t)\exp{ikz - i\omega t}\right]\label{eq:electric},
\end{align}
where $E$ is propagating in the $z$ direction with carrier frequency of $\omega$, with the wave-vector $k = \bar n\omega/c$ evaluated from  the average permittivity. 
The slow varying amplitude $A$ can also be separated into an average value and fluctuations for a stationary state, thus $A(\vb r, t) = \bar A(\vb r) + \delta A(\vb r, t)$. Evaluations of the fluctuation will consider the Fourier transform  $\delta A(\vb r, \Omega)$, which can be associated with sideband components of the central carrier at frequencies $\omega\pm \Omega$.
Moreover, from the fact that the field evolves inside a cavity, the spatial profile of the amplitude can be associated with a given spatial mode $\mu$ of the cavity, 
hereby denoted as the set $\{u_\mu(\vb r)\ |\ (\nabla^2 + k_\mu^2)u_\mu = 0\}$ in such a way that $\bar A(\vb r) = \alpha u_\mu(\vb r)$, where $\alpha$ is a complex number regarding the amplitude in a stationary regime. In what follows, we consider only a single cavity mode, thereby ignoring the other nonresonant modes.

One can show \cite{PhysRevA.79.063816} that the electromagnetic wave equation with variable permittivity can be described as
\begin{align}
    \nabla^2 \vb E  = \mu_0\bar\epsilon\pdv[2]{\vb E}{t} + \mu_0\pdv[2]{\delta\epsilon\vb E}{t}\label{eq:maxuello}.
\end{align}
The contribution in fluctuation through a small propagation distance $\delta z$ due to a fluctuation in the material permittivity can be evaluated in the spectral domain, thus for $\delta\epsilon(\vb r, \Omega)$ we have the leading contribution
\begin{align}
    \delta A(\vb r, \Omega) &= i\frac{k}{2\bar\epsilon}\bar A(\vb r)\delta\epsilon(\vb r, \Omega)\delta z\label{eq:fluctuation},
\end{align}
discarding higher-order terms. 
In this context, the noise increment is linear in the field amplitude. It is also important to notice that the fluctuations added to the field are in quadrature with the carrier amplitude, given the imaginary unity multiplying the term in \autoref{eq:fluctuation}. In that sense, it is said that the excess noise 
is added to the \emph{phase quadrature} of the involved field.

The total contribution of this fluctuation over the volume of the material to the phase quadrature fluctuation in a specific cavity mode,  namely $\delta Q(\Omega)$, is obtained from the sum of contributions of each amplitude fluctuation projected in the mode itself, plus the summation of this effect throughout the hole path inside the cavity. Using the orthonormality of the chosen set of spatial modes, one can then conclude that $\delta Q_\mu(\Omega)$ is given by
\begin{align}
    \delta Q_\mu(\Omega)&= -i\int\sum\limits_\nu\left[\iint\delta A(x,y,z, \Omega) \cdot u_\mu(x,y,z)^*\dd{x}\dd{y}\right]\dd{z}\nonumber\\
    &=\alpha \frac{k}{2\bar\epsilon}\int\limits_V \abs{u_\mu(\vb r)}^2 \delta\epsilon(\vb r, \Omega)\dd[3]{r}\label{eq:pn_added},
\end{align}
where $V$ is the whole volume of the cavity. Multiplication by $-i$ takes into account that the result explicitly addresses the contribution that is in quadrature to the mean field. This will be taken into account when we use the result in \autoref{eq:pn_added} in the contribution to the stochastic treatment for a field inside a cavity.

Describing the field by its quadratures, it is convenient to define a vector containing all the quadratures fluctuations of the cavity modes as
\begin{align}
    \va X = \bigoplus\limits_\mu (\delta p_\mu,\delta q_\mu)^\top\label{eq:X},
\end{align}
where the direct sum is taken over the set of Hilbert spaces of all cavity modes, and the real values $\delta p_\mu$ and $\delta q_\mu$ are the Wigner representation of the fluctuations of amplitude and phase quadratures of the fields in cavity mode $\mu$.

With that definition, one can obtain, using the \emph{input-output} formalism \cite{walls2008quantum}, a Langevin equation regarding the time evolution of the intra-cavity field given by
\begin{align}
    \partial_t\va X =  M_D\va X + T \va X_{in} + T'\va X_{vac} + \va Q\label{eq:time_Langevin},
\end{align}
where the time $t$ is represented in units of the cavity round-trip, $\va X_{in}$ represents the input fields, $\va X_{vac}$ the vacuum fields entering the cavity through spurious losses, the $M_D$ is the drift matrix, the $T$ and $T'$ are matrices representing the coupling of the ports of the cavity,
and $\va Q = \bigoplus\limits_\mu (0,\delta Q_\mu)^\top$ is the extra noise contribution from the permittivity fluctuation. 

In \autoref{eq:time_Langevin}, the drift matrix $M_D$ describes the evolution of the modes in a round trip, which might include phase shift and parametric gain, or attenuation. The parametric process will not be considered in the present analysis, as all measurements were made below the OPO's oscillation threshold. In that consideration, the drift matrix accounts only for the phase shift and losses and is given by
\begin{align}
 M_D = -\bigoplus\limits_\mu(t_\mu^2 + t_\mu'^2)
 \begin{bmatrix}
1 & \Delta_\mu \\
-\Delta_\mu & 1
\end{bmatrix}
\label{eq:drift},
\end{align}
where $t_\mu$ and $t'_\mu$ are the transmission coefficients for the coupling losses and spurious losses respectively of the mode $\mu$, and $\Delta_\mu$ is the detuning of that mode with respect to the cavity $\Delta_\mu = \tfrac{\omega_\mu-\omega_c}{\delta\omega}$ with $\omega_c$ being the frequency of the closest resonant mode of the cavity and $\delta\omega$ being the cavity bandwidth. Meanwhile, the coupling matrices are simply given by
\begin{align}
    T = \sqrt{2}\bigoplus\limits_\mu\operatorname{diag}(t_\mu,t_\mu)\quad \&\quad T' = \sqrt{2}\bigoplus\limits_\mu\operatorname{diag}(t'_\mu,t'_\mu)\label{eq:coupling_m}.
\end{align}
Notice that in the absence of the parametric processes, distinct field modes inside the cavity do not interact with each other, and all the matrix terms in \autoref{eq:time_Langevin} are block-diagonal. 

Assuming then a stationary regime, one can take a Fourier transform of the \autoref{eq:time_Langevin} to explore the fields by its spectral components as described in \autoref{eq:fluctuation} and \autoref{eq:pn_added}, thus leading to 
\begin{align}
    \va X(\Omega) = \left[i\Omega(T^2 + T'^2) + M_D\right]^{-1}\left(T \va X_{in}(\Omega) + T'    \va X_{vac}(\Omega) + \va Q(\Omega)\right)\label{eq:freq_Langevin}.
\end{align}

\autoref{eq:freq_Langevin} gives the dynamics of intra-cavity modes. But the actual field that one can measure is the field $\va X_{out}(\Omega)$ that is reflected from the cavity. This field can be related to the input and intra-cavity fields using the input-output formalism \cite{walls2008quantum} to get the relation between external and internal fields in
\begin{equation}
    \va X_{out}(\Omega) = T \va X(\Omega) - \va X_{in}(\Omega)\label{eq:input-output}.
\end{equation}

From \autoref{eq:freq_Langevin} and \autoref{eq:input-output}, the moments of the reflected field from the cavity can be determined. Since the state of the output fields in OPOs at the regimes of work in this research is generally Gaussian \cite{PhysRevA.92.012110}, all the information about those fields is encoded at most in the second-order moments. That is to say that by obtaining the 
spectral 
covariance matrix $V(\Omega) = \expval{\va X(\Omega)\va X(-\Omega)^\top}$ 
of a field $\va X$, all the information about it can be recovered.

Finally, the covariance of the output field $V_{out}$ is given by
\begin{align}
    V_{out}&= V'_{in} + V_{loss} + V'_{Q}\label{eq:Vout},\\
    V'_{in}&=\left[T\left[i\Omega(T^2 + T'^2) + M_D\right]^{-1}T - \mathbb{I}\right]V_{in}\left[T\left[-i\Omega(T^2 + T'^2) + M_D\right]^{-1}T - \mathbb{I}\right]^\top,\\
    V_{loss}&=T\left[i\Omega(T^2 + T'^2) + M_D\right]^{-1}T'T'\left[\left[-i\Omega(T^2 + T'^2) + M_D\right]^{-1}\right]^\top T,\\
    V'_Q &= T\left[i\Omega(T^2 + T'^2) + M_D\right]^{-1}V_Q\left[\left[-i\Omega(T^2 + T'^2) + M_D\right]^{-1}\right]^\top T,\label{eq:Vq_}
\end{align}
where $V_{in}$ is the covariance matrix of the input field and $V_Q(\Omega) = \expval{\va Q(\Omega)\va Q(-\Omega)^\top}$ is the covariance matrix of the added noise due to permittivity fluctuations.

To evaluate the extra noise added due to acoustic fluctuations, one must resort to photoelasticity and electro-optics theory \cite{narasimhamurty1981photoelastic,Wolff:21,J_E_Sipe_2016} 
associating the strain tensor $S(\va r, t)$ 
to changes in the index of refraction of a material through the 4$^{th}$ order tensor $p$ as in
\begin{align}
    \delta \epsilon_{jk} = \frac{\bar\epsilon_{jj}\bar\epsilon_{kk}}{\epsilon_0}p^{\ \ \ ql}_{jk}S_{ql}(\va r, t)\label{eq:pokels},
\end{align}
where the indices represent the three optical axes of the crystal and $\epsilon_0$ is the vacuum permittivity. In \autoref{eq:pokels}, the material dispersion was disregarded for simplicity, given the narrow bandwidth of the involved frequencies around $\omega$.

With \autoref{eq:pokels} and \autoref{eq:pn_added} one can obtain the non-null terms of $V_{Q\mu\nu}(\Omega) = \expval{\delta Q_\mu(\Omega)\delta Q_\nu(-\Omega)}$ in
\begin{equation}
    V_{Q\mu\nu}(\Omega) = \alpha_\mu\alpha_\nu^*\frac{k_\nu k_\mu}{4\bar\epsilon^2}\iint\limits_V \abs{u_\mu(\vb r)}^2 \abs{u_\nu(\vb r')}^2 p^{\ \ ql}_{jj}p^{\ \ mn}_{jj}\expval{S_{ql}(\va r, \Omega)S_{mn}(\va r', \Omega)}\dd[3]{r}\dd[3]{r'}\label{eq:Vq},
\end{equation}
already imposing an interaction of only one optical polarization, as will be discussed in \autoref{Met_and_Mes}.

The integration in \autoref{eq:Vq} contains a modulus overlap integral of the spatial modes, weighted by the correlation between the spectral components of the strain tensor associated with these modes. Assuming that the coherence length of the thermal acoustic waves is much smaller than the characteristic length of the spatial modes, the correlation between the strain tensors can be approximated as a Dirac delta function in position and mode. This approximation eliminates the convolution and simplifies the integral. Regardless of the resolution, the integral converges to a constant $c_{\mu\nu j}(\Omega)$, which is related to the spectral correlation between strain modes weighted by the optical spatial-mode overlap. Consequently, one finds that $V_{Q\mu\nu}(\Omega)$ is proportional to the product of the amplitudes of modes $\mu$ and $\nu$ \cite{PhysRevA.79.063816}, as expressed in
\begin{equation}
    V_{Q\mu\nu}(\Omega) = \eta_{\mu\nu}(\Omega)\alpha_\mu\alpha_\nu^*\label{eq:Vq2}.
\end{equation}

An important remark about \autoref{eq:Vq2} is that the extra phase noise added to a field mode is then proportional to its intensity $\Delta^2 Q = \eta(\Omega) \abs{\alpha}^2$. The noise coupling coefficient $\eta$ is related to the variance of the strain tensor, which in turn relates to the phonon population distribution \cite{J_E_Sipe_2016}. Notice that this corroborates the notion that Brillouin scattering is a non-linear process that is amplitude dependent on both optical and acoustic fields. That notion is widely employed to mitigate this excess noise by cooling the crystal lattice, thereby reducing the phonon population and consequently decreasing phase noise \cite{PhysRevA.79.063816}. With that, one can expect that the behavior of a Bose-Einstein distribution presents itself over temperature changes in the noise coupling coefficient. 

Within the temperature ranges of this research, we expect the phonon population to increase monotonically with the temperature.


\section{Experimental setup and methodology}\label{Met_and_Mes}
\begin{figure}[H]
    \centering
    \includegraphics[width=\textwidth]{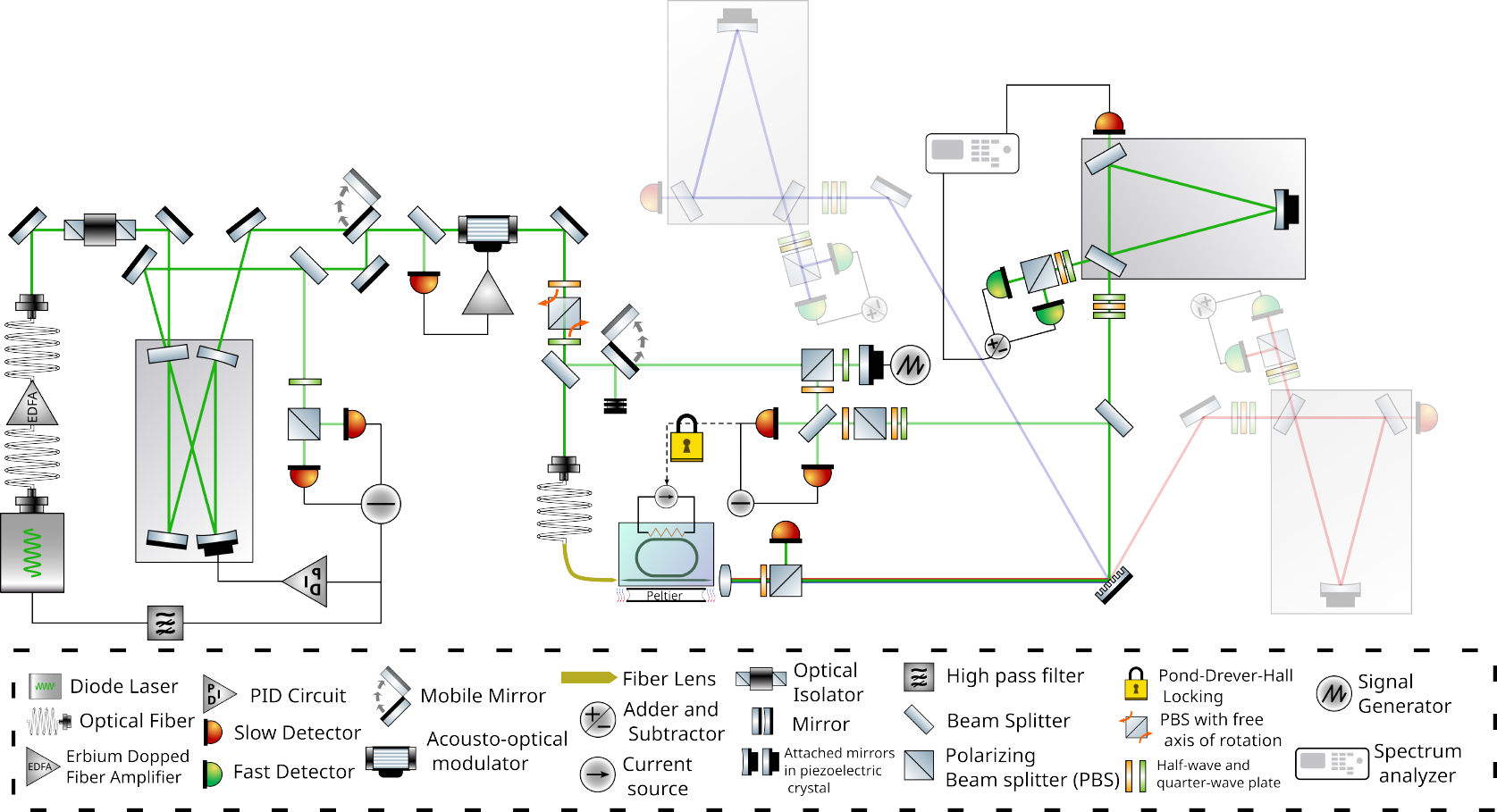}
    \caption{Diagram of the complete measurement setup. The shaded area indicates the portion of the system not utilized in this work, since all measurements are performed below the OPO's oscillation threshold and therefore do not produce a nonzero average beam pair for detection.}
    \label{fig:setup}
\end{figure}

The experiment begins with the preparation and characterization of the pump field. The micro-ring is then set to a specific temperature at which its coupling and spurious losses are determined (as described in \autoref{cav_charact}). These parameters are subsequently used with the characterization of the reflected field to infer the added noise in the intra-cavity field. All the field characterizations are made using resonant-assisted homodyne detection (RAHD) \cite{PhysRevA.88.052113,PhysRevLett.111.200402}. A diagram of the experimental setup is shown in \autoref{fig:setup}.

The pump source is a continuous-wave (CW) diode laser operating at a wavelength of 1560.599 nm, and the output is subsequently amplified by an erbium-doped fiber amplifier (EDFA). Diode lasers inherently exhibit significant phase noise \cite{10.1063/1.92254,1071522,T-CZhang_1995}, which would contaminate the detected signal. 
To mitigate this effect, a 4 m-long filter cavity with a bandwidth of 300 Hz is employed, reducing the excess noise to at most 4 dB above the standard quantum level (SQL) limit in the worst-case scenario. The cavity is locked to resonance using the Hänsch–Couillaud (HC) technique \cite{hansch1980laser}. The resulting error signal provides feedback to both a slow control that acts on a piezoelectric transducer in the cavity and a fast control that modulates the diode laser current. Residual intensity fluctuations remain after this stabilization stage; these are further suppressed by a feed-forward scheme that controls the amplitude of the driving signal of an acousto-optic modulator (AOM) operating at a central frequency of 54 MHz.

After filtering, the pump beam passes through power and polarization controls. The beam is then
inserted into an optical fiber-lens that couples light into the inverse-tapered wave guide at the entrance of the chip \cite{cardenas2014high}. The control part consists of a half-wave plate, a polarizing beam splitter (PBS), and a quarter-wave plate. With this set of optical elements, one can establish any polarization at any power below the source power at the entrance of the micro-cavity by countering the effects of the fiber and couplings' birefringences.

Polarization control is important for precisely aligning the injected field with the waveguide modes, selecting only those that are coupled to the microcavity mode of interest. This microcavity was used in \cite{Alfredo.24}, where the waveguide with 2630 nm × 730 nm cross-section couples to the microcavity that is built with a closed-loop with a free spectral range of 80 GHz.  
The cavity will accept two polarization modes, transverse electric (TE) and transverse magnetic (TM), according to the orthogonality of the electric and magnetic fields to the direction of propagation \cite{okamoto2000fundamentals}.
As noted in the literature \cite{Kakihara:06}, the TM mode experiences greater propagation loss, particularly in curved guides. Thus, for a higher-quality-factor cavity, the polarization control imposes the TE mode at the cavity entrance.

Power control is also fundamental to this research. As stated before in \autoref{Mat_Frame}, the phonon noise presents itself as proportional to the optical field power, as described in \autoref{eq:Vq2}. To experimentally obtain the coupling coefficient $\eta$, a linear fit of the noise over a range of pump powers must be taken. That entails that the pump noise must be known for each input power on the cavity so that the $V_{in}$ term in \autoref{eq:Vout} can be used to infer the desired $V_Q$ out of the measurement of $V_{out}$ for each intensity.



The temperature of the waveguides strongly determines the effective optical path length of the Si$_3$N$_4$ micro-ring through the thermo-refractive effect. This implies that, for a fixed wavelength, there is a discrete 
set of temperatures at which the cavity resonates.
To select and control the resonant conditions, the experimental setup incorporates two temperature-control mechanisms. The first is a slow, global temperature control provided by a Peltier plate mounted beneath the chip's copper base, operating over a range from $20^\circ$ C up to $140^\circ$ C. The second is a fast local temperature control implemented using a platinum resistance embedded on a section of the cavity waveguide, enabling localized heating. Using this configuration, the resonant condition is initially selected with the Peltier element, and the cavity is subsequently locked to resonance by feeding back the transmission signal to the platinum heater via a Dither-and-Lock technique at a 10 kHz modulation frequency.

It is worth noting that, within the safe current range, one can achieve two distinct resonant conditions, here denoted $R1$ and $R2$, at the same basal chip temperature. The Peltier temperature is selected to minimize the $R1$ current signal while maintaining sufficient current for the control driver to operate. 
Acquisitions were performed at successive chip temperatures for both available resonances. Notice that the resonance $R_2$ at temperature $T_1$ is the same resonance $R_1$ for a subsequent temperature $T_2>T_1$. The main text will focus on the analysis with the resonance at lower current $R_1$, while the corresponding results at $R_2$ are given in the appendix.

To characterize the fields in the aforementioned configurations and reconstruct the covariance matrix $V_{out}$ in \autoref{eq:Vout}, a resonant-assisted heterodyne detection (RAHD) scheme is used \cite{PhysRevA.88.052113,PhysRevLett.111.200402}. In traditional heterodyne detection (HD), a weak field of interest is interfered with an intense, known field, the local oscillator (LO), at a beam splitter. The observed photocurrent will present fluctuations associated with the beatnote of the frequencies of interest with the local oscillator frequency. An electronic local oscillator is then used to recover the terms of the sidebands under analysis and infer the statistical moments of combinations of field modes at a selected spectral distance from the LO. 
The RAHD method shares the same core idea as the HD, but the key distinction is that the relatively strong carrier of the field is used as the local oscillator for the weak sidebands.
The key distinction is that field-phase control is implemented by a dispersive component, such as a cavity, which can selectively apply a phase shift to the carrier and each sideband. One advantage of this method is its application to reconstruct the covariance matrix from intense fields, which would not be possible with the usual HD technique.
Another advantage of the RAHD technique is its ability to distinguish the role of each sideband, thus spanning an extended information by accessing two modes of the field \cite{PhysRevLett.111.200402}. 
In this project, the characteristic of phase fluctuations involved allows us to assume sideband symmetry, namely that the same process populates both sides, and thus, we consider only a one-dimensional projection of each accessible subspace at each time for simplicity.  In the current setup (\autoref{fig:setup}), the detection is composed of a cavity, with a Free Spectral Range of 700 MHz and a bandwidth of 4 MHz, and a balanced detector, with the photocurrents that can be added or subtracted. The shaded regions are the implementations used in \cite{Alfredo.24} that are not used in the current experiment.


\begin{figure}[tbh!]
    \centering
    \includegraphics[width=\linewidth]{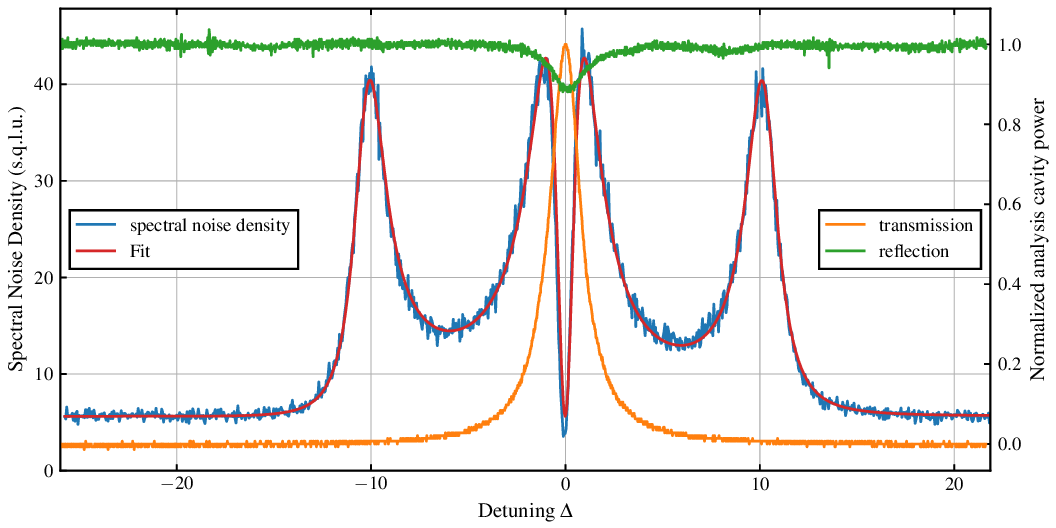}
    \caption{Spectral noise for a scanning detection cavity. The transmitted signal is used to calibrate the detuning, normalized here by the half bandwidth at half maximum. Noise power has the electronic noise subtracted and is normalized to the standard quantum limit by the shot noise level measurement. The continuous line is the adjustment of \autoref{eq:SR}. Analysis frequency of 20 MHz, cavity bandwidth of 4 MHz.}
    \label{fig:elipsisroll}
\end{figure}


The spectral noise density of the photocurrent signal, in standard quantum level units (s.q.l.u.), follows \cite{PhysRevA.88.052113}:
\begin{align}
	SN(\Delta,\Omega)&=c_{\alpha}(\Delta)\alpha + c_{\beta}(\Delta)\beta + c_{\gamma}(\Delta)\gamma +c_{\delta}(\Delta)\delta + c_{vac}(\Delta),\label{eq:SR}\\
    &\begin{aligned}
        c_{\alpha}(\Delta)= \abs{g_+}^2,\ &\  c_{\beta}(\Delta)=\abs{g_-}^2,\\
    c_{\gamma}(\Delta)=2\operatorname{Re}(g_+g_-^*),\ &\  c_{\delta}(\Delta)=2\operatorname{Im}(g_+g_-^*),\\
    c_{vac}(\Delta)=1-&\abs{g_+}^2-\abs{g_-}^2,
    \end{aligned}\nonumber
\end{align}
where $\Delta$ is the analysis resonator detuning with respect to the signal carrier
normalized by the cavity half bandwidth, 
$g_+$ and $g_-$ are linearly independent functions of the detuning $\Delta$  and the frequency of analysis $\Omega$ given by
\begin{align}
    r(\Delta) &= \frac{r_m-i\Delta}{1-i\Delta},\quad R(\Delta,\Omega) = \frac{r(\Delta)^*}{\abs{r(\Delta)}}r(\Delta + \tfrac{2\Omega}{\operatorname{BWD}}),\nonumber\\
    g_+ &= \frac{R(\Delta,\Omega)+R(\Delta,-\Omega)^*}{2}\quad g_- = i\frac{R(\Delta,\Omega)-R(\Delta,-\Omega)^*}{2}\label{eq:gs}
\end{align}
where BWD is the analysis cavity bandwidth and $r_m$ is its reflectivity at resonance.

The $\alpha$, $\beta$, $\gamma$ and $\delta$ coefficients compose the spectral matrix 
$V$, evaluated in \autoref{eq:Vout}, as 
defined in \autoref{eq:spectral_M}. The $\alpha$ term gives the variance in the amplitude quadratures in relative basis, $\beta$ the phase quadratures, $\gamma$ the covariance between phase and amplitude quadratures, and $\delta$ the imbalance between upper and lower sidebands (which was assumed to be zero as mentioned previously). 
\begin{equation}
    V = \begin{pmatrix}
        \alpha&\gamma+i\delta\\
        \gamma-i\delta&\beta
    \end{pmatrix}\label{eq:spectral_M}.
\end{equation}

The measurement of the spectral noise is performed by a spectrum analyzer during the scanning of the analysis cavity.
as shown in \autoref{fig:elipsisroll}. The spectrum analyzer is configured to acquire 1001 points over a 50 ms sweep, with a 20 kHz video bandwidth and a 100 kHz resolution bandwidth. Giving the synchronization between the cavity scanning and the spectrum analyser acquisition, 
each noise acquisition point is related to a single detuning point, so one can take as many averages of the spectral noise density as desired. In this experiment, each noise measurement is the average of 100 acquisitions. 
In \autoref{fig:elipsisroll}, we present the result for an acquisition at an analysis frequency of 20 MHz. The weak transmission of the cavity is recorded, and its Lorentzian profile gives the reference for the detuning, normalized here by the cavity half-bandwidth. It matched the reflected intensity, which shows a weak depletion  (10 \%) on resonance. 
Residual electronic noise is subtracted, and the shot noise level is obtained from the same measurement using the noise power of the subtraction of the photocurrents for normalization purposes.
The results are then adjusted to the model given by \autoref{eq:SR}, with the assumption $\delta = 0$. As an example, the result in \autoref{fig:elipsisroll} 
returned $\alpha = 5.666(13)$ s.q.l.u., $\beta = 43.208(29)$ s.q.l.u., and $\gamma = -0.919(17)$ s.q.l.u., and these parameters of this data set are represented by the red points (the ones with higher power) in \autoref{fig:R1_examp}. Upon the scanning of the cavity, the measured noise in the photocurrent shifts from the amplitude to phase quadrature measurement of the incoming field when the cavity is resonant with the sidebands, or when it is half the resonance peak \cite{10.1119/1.2937903}.

The measurement recovers the amplitude and phase quadrature variances of the field and their correlation, and the experiment is repeated for distinct pump powers, leading to the results presented in \autoref{fig:R1_examp}. The procedure is repeated for different temperatures, for resonances $R_1$ and $R_2$ in each temperature, for distinct couplings between the microcavity and the waveguide. Yet, the evaluation of the intracavity noise added at each implementation depends dramatically on the cavity parameters (\autoref{eq:coupling_m}). The determination of these parameters is discussed next.


\section{Micro-ring characterization}\label{cav_charact}

\begin{figure}[tbh!]
    \centering
    \includegraphics[width=\linewidth]{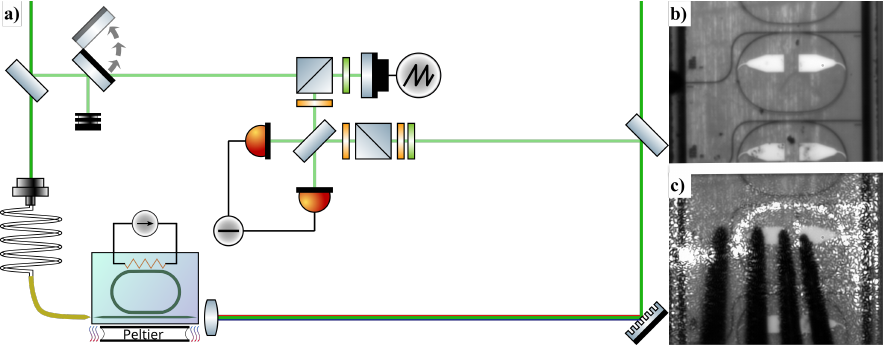}
    \caption{Diagram of the part of the setup that constitutes the Mach-Zehnder interferometer (a) and pictures of the chip taken with a near-infrared camera. Micro-ring (b) has a height of 390 $\mu$m and a width of 540 $\mu$m. The white pads are contacts of the platinum wire. Once the contacts run a current, the cavity can be tuned into resonance (c), and some scattering from the surface of the waveguide can be observed. The interaction length between the bus waveguide and the cavity waveguide is about 200 $\mu$m.}
    \label{fig:mach_zehnder_chip}
\end{figure}

The experiment is performed in four cavities with the same racetrack ring geometry. 
The only geometrical difference between these microcavities is the size of the evanescent coupling gap, and thus, in theory, they should differ only in coupling loss. We used three cavities from the same chip with gaps of 250 nm, 275 nm, and 300 nm; from another identical chip, a cavity with a 250 nm gap was also used for consistency. All waveguides are made of amorphous Si$_3$N$_4$ 
and are embedded in a pure SiO$_2$ substrate. 
The micro-cavity geometry can be seen in \autoref{fig:mach_zehnder_chip}b,c. Given the cavity perimeter of roughly 1.5 mm, the Free Spectral Range of 80 
GHz is common for all microcavities.

Coupling losses and spurious losses depend on a lot of instances \cite{okamoto2000fundamentals} such as the index of refraction that, due to the thermo-refractive index, depends on the temperature of the waveguides. The confinement of the waveguides depends on the modulus of the difference in the index of refraction of the waveguide core and substrate. 
Coupling strength between the waveguides will depend on this confinement, as well as on the interaction length.
At the same time, the effective interaction length also depends on the waveguides' index of refraction, which means that the back-and-forth exchange of energy \cite{okamoto2000fundamentals} between waveguides is affected by temperature as well. The propagation losses are also affected by the waveguide confinement, as more confined waveguides have smaller losses for the same propagation distance. Therefore, to determine $V_Q$, one must at each temperature determine the $T$ and $T'$ coefficients.

To determine the coupling and spurious losses of a micro-ring resonator, the standard procedures used for macroscopic cavities, in which the losses of individual, mountable components can be measured independently, are not applicable. Nevertheless, a laser frequency scan can still be used, allowing to fit the reflected field with a Lorentzian function and extract the cavity half-bandwidth and the minimum reflectance, which occurs when the cavity is on resonance. In our setup, this is achieved by bypassing the filter cavity with a movable mirror (see \autoref{fig:setup}) and by scanning the diode laser frequency by varying the driving current. Together with the free spectral range, these quantities determine the cavity finesse. From this information, the two loss parameters can be inferred, but only up to an exchange symmetry between coupling and spurious losses. 

\begin{figure}[tbh!]
    \centering
    \includegraphics[width=\linewidth]{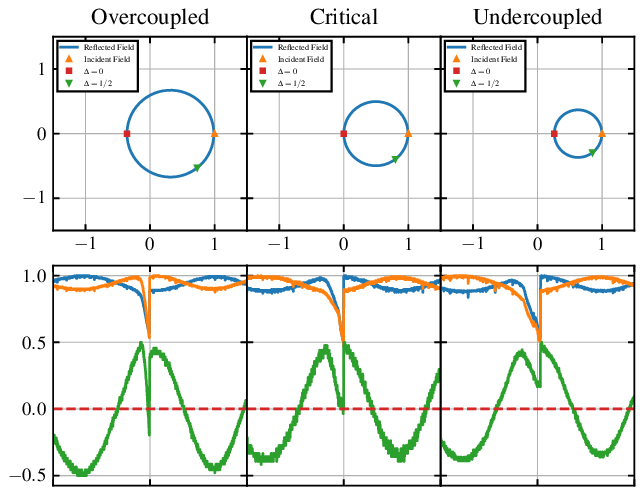}
    \caption{Graphs of the reflected amplitude for the different cavity coupling regimes. In the graphs on the top row, you can see the complex amplitude response of the reflected field in each coupling regime, where the y-axis represents the imaginary component and the x-axis the real component. In the graphs in the bottom row, the actual characterizations of the coupling regime show phase inversions in the overcoupled case when depletion crosses the dashed red line. The blue and orange curves represent the normalized detector signals, while the green curve shows their difference, rescaled for visibility. These coupling regimes were all made in the same cavity, which is typically supercritical. For the other cases, spurious losses were artificially introduced into the system by increasing the pumping power above the threshold, thus causing conversion losses for the pump field to the beam pair.}
    \label{fig:couplings}
\end{figure}

This ambiguity can be resolved using methods reported in the literature that allow identification of the resonator's coupling regime \cite{Dumeige:08,TREBAOL2009964,gao2022probing}. By determining whether the cavity operates in the under-coupled or over-coupled regime—i.e., whether the spurious (intrinsic) losses or the coupling losses dominate—it is then possible to unambiguously assign each extracted value to its corresponding loss mechanism. In general, those methods are cumbersome for repeated characterizations, as they either involve fast scans and complex aftermath \cite{Dumeige:08, TREBAOL2009964}, or scanning multiple cavities with the same geometry in the same conditions, for different coupling losses, to discriminate if the chosen cavity is below or above the critical coupling. In fact, the reason 
for the production of 
multiple cavities with the same race-track geometry is that this was the initial method used to characterize the micro-cavities.

\begin{figure}[tbh!]
    \centering
    \includegraphics[width=0.9\linewidth]{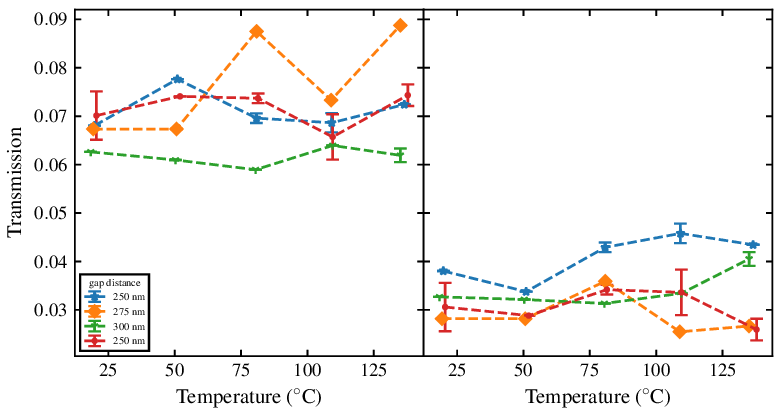}
    \caption{Graphs of the estimated coupling and spurious losses. 
    The First column presents the coupling losses, and the second column presents the spurious losses. As one can see by the graphs, in all cases the coupling regimes were overcoupled since the coupling losses were always bigger than the spurious ones.}
    \label{fig:losses}
\end{figure}

We proposed a faster method for this system to discriminate between coupling regimes at each temperature and condition. This method is based on the observation that only the overcoupled regime can invert the phase of the incident field, as shown in the first row of graphs in \autoref{fig:couplings}. This phase inversion can be readily detected using an interferometer in which one path interacts with the micro-ring, while the other passes through a predictable phase control. In the case of this experiment, we used a Mach-Zehnder interferometer, as can be seen in \autoref{fig:mach_zehnder_chip}. One can show that the difference in intensity of the detectors in the case of the balanced interferometer is proportional to the expression given by
\begin{align}
    D&\propto\frac{r(1+r'^2)\cos\delta - r'(\cos(\delta-\phi)+r^2\cos(\delta+\phi))}{1 + r^2r'^2-2rr'\cos\phi}\label{eq:diff},
\end{align}
where $r = \sqrt{1-t^2}$ and $r' = \sqrt{1-t'^2}$, $\delta$ is the path phase difference of the interferometer alone, and $\phi$ is the added phase by the cavity. It is important to notice that \autoref{eq:diff} is not invariant under $t$ and $t'$ swapping, which means it can be used to discriminate between the two losses.

In the case where the cavity is at resonance, and the path phase difference is zero, the expression in \autoref{eq:diff} simplifies to
\begin{align}
    D&\propto\frac{r-r'}{1-rr'}\label{eq:diff2}.
\end{align}
Since $0<r'<1$ and $0<r<1$, the signal of the difference only inverts in the case where $r' > r\rightarrow t'<t$, which means that the interferometer signal only inverts in the regime where the coupling losses dominate, that is, the overcoupled regime. To determine the coupling regime, one synchronizes the micro-ring scan produced by its embedded heater with the interferometer's piezo scan to match the resonance condition to a peak in the interferometer's response, enabling detection of signal inversion. With that, the losses obtained from the Lorentzian fit are then discriminated. The second row of \autoref{fig:couplings} presents examples of this characterization. In order to emulate distinct losses on the ring, we operate the same microcavity under different pump regimes. In the first one, since $D<0$, coupling losses dominate. In order to have higher losses, we increased the pump power, reaching the oscillation threshold of the optical parametric oscillator for demonstration purposes only. We could achieve either a perfect match $D\simeq0$ associated with critical coupling, or $D>0$ in a situation of undercoupling.

The analysis was performed in four rings under distinct temperatures for both resonance conditions $R1$ and $R2$. 
The graphs in \autoref{fig:losses} present the results for resonance $R1$, and demonstrate that, while the rings are always overcoupled, coefficients may present great variations with the temperature, which must be carefully taken into account in the evaluation of the added noise inside the resonator.


\section{Results and analysis}

\begin{figure}[h!]
    \centering
    \includegraphics[width=\linewidth]{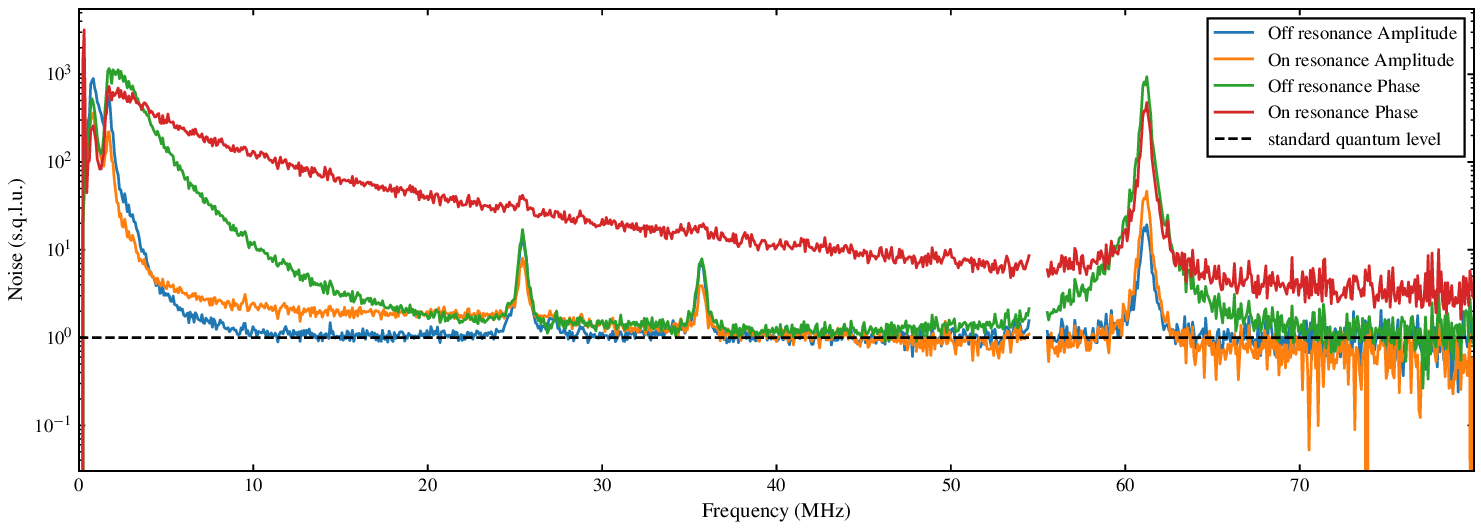}
    \caption{Normalized noise power as a function of the analysis frequency. 
    Spectrum analyzer parameters are 200 kHz of resolution bandwidth, 20 kHz of video bandwidth, 50 ms of sweep time, and 2000 acquisition points. Phase and amplitude measurements were performed with the help of the analysis cavity. 
    }
    \label{fig:spectrum_scan}
\end{figure}

Characterization of the noise begins with the spectral analysis of the output of a resonating microcavity. The goal is the observation of the spectral region with added noise from the microcavity itself and spurious sources that may affect the measurement.
\autoref{fig:spectrum_scan} presents amplitude and phase quadrature noise power (normalized to the vacuum level), with the microcavity in and out of resonance. Looking for the results for the cavity out of resonance (blue and green), we can notice that amplitude fluctuations of the laser, transmitted through the embedded waveguide, are shot noise limited above 10 MHz, and phase fluctuations are reduced to the coherent state level at 40 MHz. While not completely suppressed, they are limited above 10 MHz, thus allowing the characterization of the microcavities above this frequency. Nevertheless, intrinsic laser noise is presented at 26, 36, and 61 MHz as broad peaks, and these frequencies were avoided. A narrow peak at 54 MHz, associated with the driving signal of the AOM, is suppressed in the graph. 

Once the microcavity, with a bandwidth of 80 MHz, is brought into resonance, the contribution from the intrinsic noise is present. We observe a clear increase in the amplitude noise below 20 MHz, and a broad and monotonically decreasing noise in the phase. This additional phase noise extends up to 80 MHz, reaching the detection limit of our electronics. Given this profile, we have chosen the analysis frequency of 20 MHz as a good point, matching the responsivity of our electronics with a clear signature of the added phase noise.



\begin{figure}[h!]
    \centering
    \includegraphics[width=0.49\linewidth]{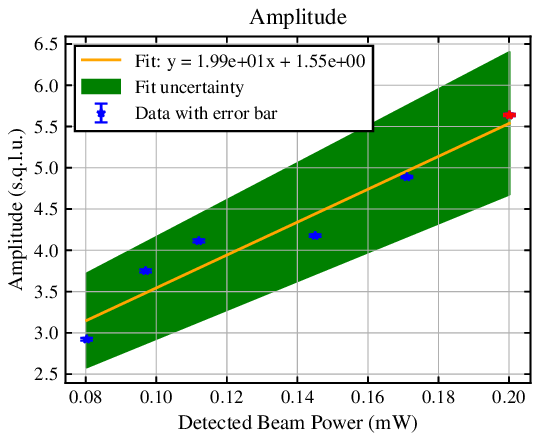}
    \includegraphics[width=0.49\linewidth]{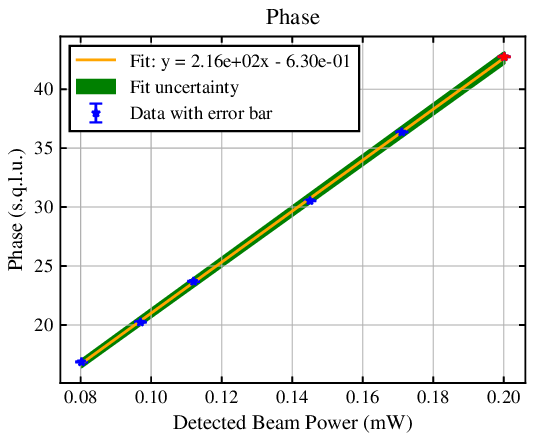}
    \includegraphics[width=0.49\linewidth]{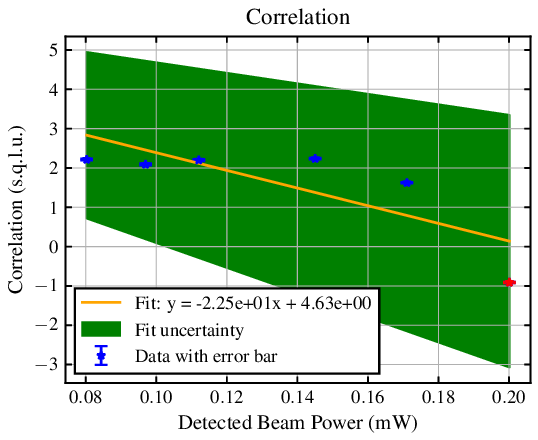}
    \caption{Graphs of an example of the linear fits of the reflected resonant field characterization parameters. These measurements were performed at a basal temperature of 18.37 $^\circ$C under resonance condition $R1$. The first two graphs present the $\alpha$ and $\beta$ coefficients for a range of powers from 0.05 mW to 0.2 mW, and the last graph presents $\gamma$. Each data point corresponds to a parameter obtained from a fit performed as in the graph in \autoref{fig:elipsisroll}; the red point (obtained at higher power) corresponds to the data presented there.}
    \label{fig:R1_examp}
\end{figure}

Previously to the microcavity analysis, the laser field characterization, using the technique presented in the \autoref{fig:elipsisroll}, is repeated for distinct powers.
The pump characterization shows a field that maintains its amplitude noise $\alpha$ at the standard quantum level, exhibits no significant covariance $\gamma$ between its quadratures (as expected for a vacuum state), and exhibits a linear increase of 0.6 s.q.l.u./mW in the phase quadrature. 
This response comes from the original phase noise of the laser, which is linearly reduced by the power control at the input of the chip. 
However, since the filtered source noise is low, the worst-case scenario does not exceed 4 dB above the standard quantum level. The measurements in the micro-cavities also do not reach the source's maximum power, as we are operating well below the OPO's oscillation threshold. Typically, the maximum input power used in resonant field characterization is approximately 0.5 mW, corresponding to 1.3 s.q.l.u. of phase noise in the input field. 

\begin{figure}[tbh!]
    \centering
    \includegraphics[width=\linewidth]{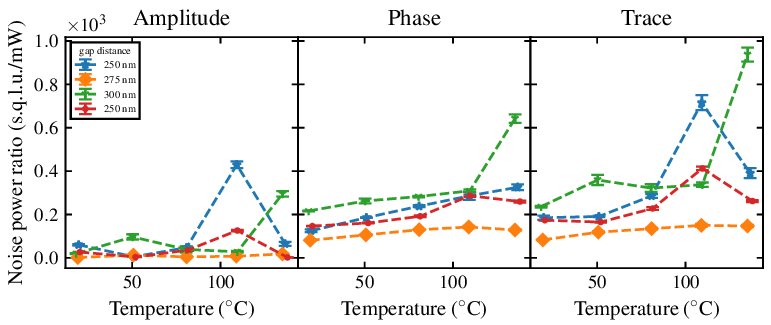}
    \caption{Graphs of the measurements of the spectral matrix components for four micro-cavities, under the resonance $R1$ condition. 
    Each traced line represents the results for a single micro-ring.}
    \label{fig:measures}
\end{figure}

With the micro-cavity in resonance, the same method of characterization of the noise for increasing laser power is applied to the reflected field to get results like the data in the graph of \autoref{fig:R1_examp}. We can see the linear increase in phase noise with field power as predicted by \autoref{eq:Vq2} in consonance with measurements for phonon noise \cite{PhysRevA.79.063816}.
This linear behavior is shown alongside a linear fit, yielding the \textbf{noise power ratio} (see \autoref{fig:measures}), which will be used to determine $\eta$. Notice, however, that the amplitude noise presents an increase as well, although with a smaller slope.
The reason for this problem is the imperfect locking of the PDH, as the competition of the thermal self-phase modulation tends to keep a small detuning for the microcavity. As we see in \autoref{eq:drift}, this detuning $\Delta_\mu$ couples the quadratures, and part of the phase noise is transferred to the amplitude.

\begin{figure}[tbh!]
    \centering
    \includegraphics[width=\linewidth]{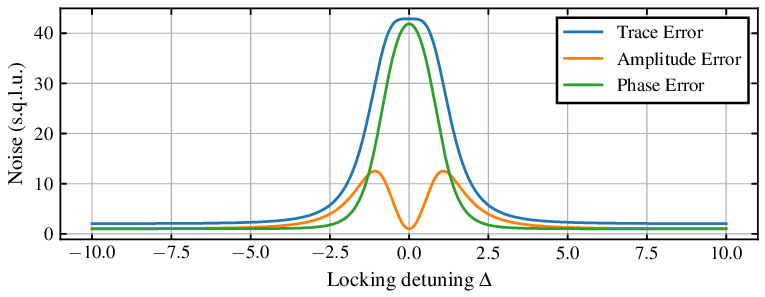}
    \includegraphics[width=\linewidth]{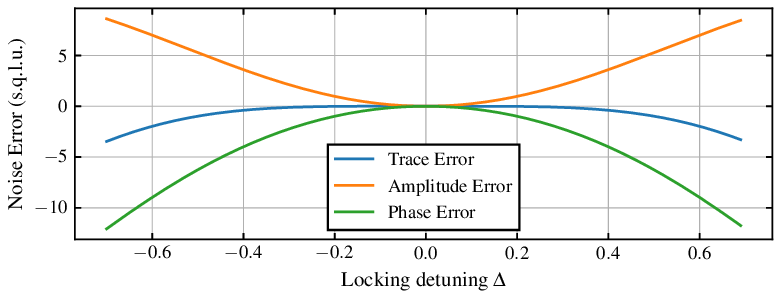}
    \caption{Graph of the error estimation for measurements of the amplitude noise, phase noise, and trace of the spectral matrix with respect to the locking error of the micro-ring. The curves in the second graph were obtained by the difference between the measurable field with the cavity near resonance and the cavity at exact resonance. For this graphical representation $\alpha = 1$, $\beta = 1$, and $\gamma = 0 $ in a cavity with coupling loss of $t = 0.62$ spurious loss of $t' = 0.032$, and $\eta = 0.0003$. Those parameter values simulate the data in the graph in \autoref{fig:elipsisroll}, but with a perfectly coherent input field.}
    \label{fig:lock_error}
\end{figure}

Since that is not a predictable condition, and thus cannot be compensated for in the post-data processing, a numerical error estimate was taken using \autoref{eq:Vout} to analyze its consequences. We evaluated the deviation on the noise observed at exact resonance, for $\alpha = 1$, $\beta = 20$, and $\gamma = 0 $ in a cavity with coupling loss of $t = 0.063$ and spurious loss of $t' = 0.089$, as is the case for most of the measurements. 
The result, shown in  \autoref{fig:lock_error}, shows that amplitude and phase noise change quadratically in a compensating way, for a range of locking error of the order of 10\% of the cavity bandwidth. Thus, their sum, corresponding to the trace of the covariance matrix, gives a value close to the resonant condition, with a deviation of the order of 1\%. 
That is the case for all field and cavity configurations because the trace of the spectral matrix (and the covariance matrix) is rotational invariant in the Hilbert space; the only contribution to the trace error is the attenuation of the cavity, which grows more slowly than the effects of the rotation. A better estimate of the added noise is obtained from the trace, thereby minimizing experimental errors.



The acquisitions for the power noise ratio for amplitude quadrature, phase quadrature, and trace detected outside the micro-cavities are summarized in \autoref{fig:measures}. As discussed, the data for different cavities do not superimpose since the intrinsic losses are not the same, in such a way that not even the intra-cavity power in each situation preserves the same proportion to the detected power. Moreover, over the temperature range, the coupling changes in a noticeable way (\autoref{fig:losses}). Nevertheless, while the amplitude noise power ratio shows strong variations, the phase noise power ratio shows a more monotonically increasing behavior. If you look at the trace, this monotonic increasing is clearer, yet distinct, even for cavities with the same gap.

\begin{figure}[tbh!]
    \centering
    \includegraphics[width=\linewidth]{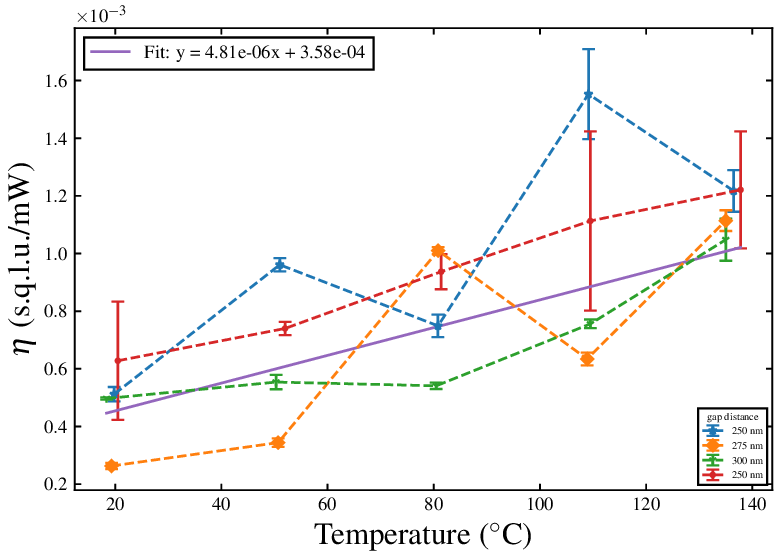}
    \caption{Graphs of the noise power ratio of the inferred intra-cavity fields for each measured temperature, for the $R1$ condition. Each traced line represents the results for a single micro-ring, and the fit considers the ensemble of the values for all the rings.}
    \label{fig:final}
\end{figure}

However, we must remember that the relevant parameter is not in the noise in the output of the cavity itself, but the coupling term $\eta$ (\autoref{eq:Vq}), evaluated from the added noise from the intracavity. From the normalized noise power that is added to the measured field on the detection system, presented in the Trace data on \autoref{fig:measures}, and the couplings, presented in \autoref{fig:losses}, we obtain the final result of the added noise per unit of power inside the cavity.
The inferred intracavity noise coupling coefficient $\eta$  for distinct microcavities presents a reasonable overlap within experimental error, as can be seen in  \autoref{fig:final}. This is consistent with our model for phonon noise contribution and our methods for its characterization.
Among the factors that define this coefficient, we have the geometry, which defines the spatial integration in \autoref{eq:Vq} on the overlap of optical modes and acoustic modes, and on the material, regarding the amplitude of the strain tensor. 
Temperature dependence is compatible with a monotonic increase, and considering the dispersion of the values, we can estimate a compatible linear response within the temperature range that is consistent with 
$\tfrac{\delta\eta}{\delta T} = 5.97(12)\times10^{-3}$ s.q.l.u./W$^\circ$C for condition $R1$.

Beyond geometric considerations, the monotonic overall behavior under temperature changes after cavity-selection compensation also provides evidence that the extra noise measured arises from electro-acoustic coupling between the intense fields inside the cavity and the acoustic modes of the waveguide's crystal lattice. As discussed in \autoref{Mat_Frame}, the non-linear interaction of the optical and acoustic fields is amplitude dependent in both fields, and temperature increase implies an increase in the thermal state phonon population.

\section{Conclusion}

 We found experimental evidence (vide \autoref{fig:final}) to support the assumption that the excess phase noise that disrupts entanglement generated by non-linear processes inside Si$_3$N$_4$ high-quality-factor micro-resonators \cite{Alfredo.24} is of phononic nature. In the process, we developed an easier method to determine the intra-cavity parameters (coupling and losses) of a resonator.
 With that tool, we conducted a complete characterization that allows us to infer the additional noise the beams
 will experience when standing in the intra-cavity environment.

Therefore, in order to achieve entanglement generation in above threshold operation, we should consider 
methods to cool the vibrations to reduce intra-cavity field scattering, thereby providing a better preservation of the information contained in those fields. This could be achieved either by direct cooling of the substrate, or by cooperative optical techniques, as shown in \cite{ConnorSkehan:23}. Another possibility is to compensate for the additional phase noise in the post-data-processing by subtracting the pump noise from the generated pair, since all internal electromagnetic fields couple to the same acoustic reservoir and thus share the contamination from the same noise source.

\begin{backmatter}
\bmsection{Funding} This project was funded by the agency Fundação de Amparo à Pesquisa do Estado de São Paulo (FAPESP) through the grants 2022/15413-8 for Yuri Sacha's PhD program, 2021/04829-6 for Gabriel's master's program, and 2022/09436-5 for the Laboratory for Coherent Manipulation of Atoms and Light (LMCAL) group funding.

\bmsection{Acknowledgment} We thank the agency Fundação de Amparo à Pesquisa do Estado de São Paulo (FAPESP) for funding this project through the grants 2022/15413-8 for Yuri Sacha's PhD program, 2021/04829-6 for Gabriel's master's program, and 2022/09436-5 for the Laboratory for Coherent Manipulation of Atoms and Light (LMCAL) group funding. We also thank 
the Nanophotonics Groups from Columbia University for their collaboration in producing the chips in the Cornell NanoScale Science and Technology Facility (CNF) labs.

\bmsection{Disclosures}
The authors declare no conflicts of interest.

\bmsection{Data Availability Statement} Data underlying the results presented in this paper are available accordingly to the data policy of FAPESP at the link
\href{https://drive.google.com/drive/folders/1fnQgBBmfojs6zFhigZUa4gCRxp4Yv90M?usp=sharing}{https://drive.google.com/drive/folders/1fnQgBBmfojs6zFhigZUa4gCRxp4Yv90M?usp=sharing}.


\end{backmatter}




\bibliography{sample}

\newpage 

\section{Appendix}

As mentioned in the main text, for a safe range of current on the platinum heater, we can access no more than two resonances of the cavity, at a chosen baseplate temperature. Thus, for safety, we had set the acquisition at two distinct currents, which are subject to small variation in order to keep the microcavity in resonance with the incoming laser.

\begin{figure}[tbh!]
    \centering
    \includegraphics[width=0.9\linewidth]{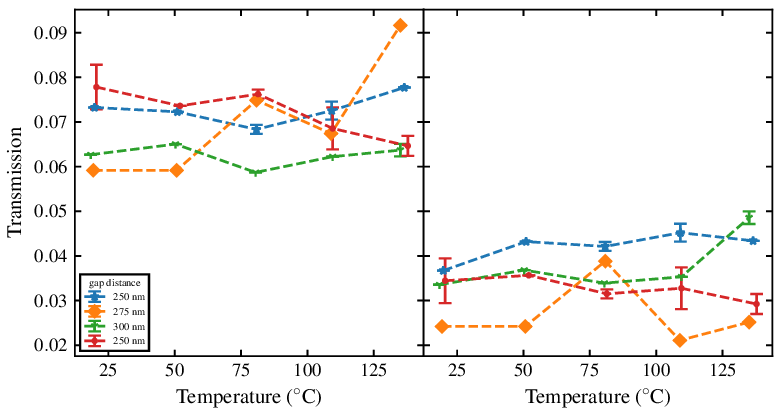}
    \caption{Graphs of the estimated coupling and spurious losses. The first row refers to the $R1$ resonance condition, while the second refers to the $R2$ condition. The First column presents the coupling losses, and the second column presents the spurious losses. As one can see by the graphs, in all cases the coupling regimes were overcoupled since the coupling losses were always bigger than the spurious ones.}
    \label{fig:losses_R2}
\end{figure}

\begin{figure}[tbh!]
    \centering
    \includegraphics[width=\linewidth]{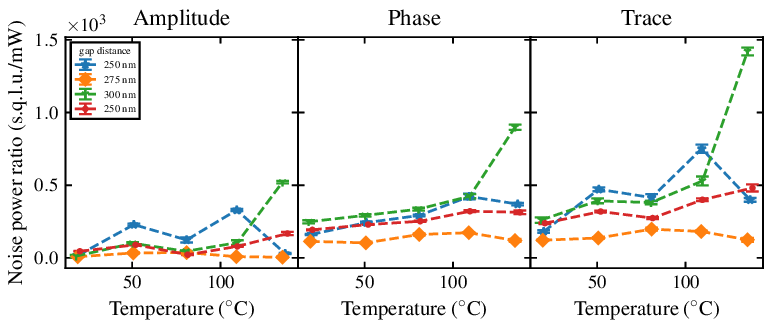}
    \caption{Graphs of the measurements of the spectral matrix components for four micro-cavities. The first row presents the results for the $R1$ condition, and the second row presents the results for the $R2$ condition. Each traced line represents the results for a single micro-ring.}
    \label{fig:measures_R2}
\end{figure}

While the first resonance was chosen for the main part of the acquisitions, we performed the characterization at the second resonance. Oddly, the obtained values for the cavity coupling and losses (\autoref{fig:losses_R2}) don't necessarily match those from the first resonance, at a higher baseplate temperature. Therefore, it is not surprising that the added noise will not present a corresponding match, as can be seen in \autoref{fig:measures_R2}.

That can be a consequence of the nonuniform temperature distribution, since the heater covers only the lower half of the ring, as can be seen in \autoref{fig:mach_zehnder_chip}. This temperature gradient may affect the resonator losses for the distinct resonant conditions $R1$ and $R2$ differently. Nevertheless, applying the same treatment used in the main text, we still can see a coupling coefficient (\autoref{fig:measures_R2}) that is compatible with a monotonic increase with temperature. The general linear adjustment gives a value of  $\tfrac{\delta\eta}{\delta T} = 6.6(22)\times10^{-3}$ s.q.l.u./W$^\circ$C for condition $R2$ that is compatible with the value obtained for condition $R1$.

\begin{figure}[tbh!]
    \centering
    \includegraphics[width=\linewidth]{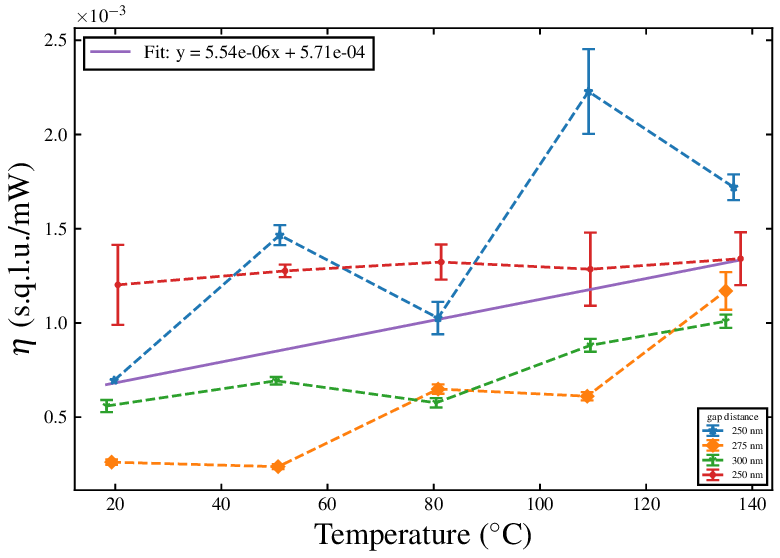}
    \caption{Graphs of the noise power ratio of the inferred intra-cavity fields for each measured temperature. The first graph shows the results for the $R1$ condition, while the second shows the results for the $R2$ condition. Each traced line represents the results for a single micro-ring.}
    \label{fig:final_R2}
\end{figure}

\end{document}